\documentclass[pra,aps,twocolumn,groupedaddress,10pt,showpacs]{revtex4-1}

\usepackage{graphicx} 
\usepackage{amsmath}
\usepackage{bbold}
\DeclareMathAlphabet{\mathbbmsl}{U}{bbm}{m}{sl}

\begin{document}
\title{On single qubit quantum process tomography for trace-preserving and nontrace-preserving maps}
\author{Ramesh Bhandari}\thanks{Email: rbhandari@lps.umd.edu}
\affiliation{Laboratory for Physical Sciences, 8050 Greenmead Drive, College Park, Maryland 20740, USA}
\author{Nicholas A. Peters}\thanks{Present address:  Computational Sciences and Engineering Division,  Oak Ridge National Laboratory, Oak Ridge, Tennessee 37831, USA.}
\address{Applied Communication Sciences, 331 Newman Springs Road, Red Bank, New Jersey 07701, USA}
\date{\today}

\begin{abstract}
We review single-qubit quantum process tomography for trace-preserving and nontrace-preserving processes, and derive  explicit forms of the general constraints for fitting experimental data.  These forms provide additional insight into the structure of the process matrix as well as reveal a tighter bound on the trace of a nontrace-preserving process than has been previously stated. We also describe, for completeness,  how to incorporate measured imperfect input states.
\end{abstract}


\maketitle


Quantum process tomography (QPT) is the gold standard method to measure and quantify  how a quantum device-under-test (DUT), such as a quantum gate or a transmission channel, will transform some arbitrary input quantum state~\cite{1,NC}.   Often a quantum process is called a map, as it maps any input state to an output state reflecting the input state's interaction with the DUT.  In this note focused on single qubit QPT, we review the treatment of a trace-preserving process (e.g., a wave-plate) and the treatment of a nontrace-preserving process (i.e., there are qubit loss or leakage errors, e.g., the impact of a polarizer on optical qubits may cause qubits to be absorbed), and extract explicit, implementable expressions from the general  constraints given in ~\cite{1,NC}. These explicit forms, which involve  different elements of the process matrix,  then provide insight into the structure of the process matrix, and thus can  serve as useful tools for an experimentalist interested in measuring quantum gates with error models including qubit leakage.  A tighter bound on the trace of a nontrace-preserving process also emerges, which we  point out.

Furthermore, we discuss input quantum states that are not perfectly prepared, which is important to consider to reduce experimental errors in the determined process matrix. Consequently, we provide  a slight modification of the  methodology for single qubit tomography ~\cite{NC} to include such imperfectly prepared input states.  To carry out single qubit QPT experimentally, one must prepare four single-qubit states which will be used to probe the DUT.  Even if states are prepared with high fidelity, they are not perfect, and as such depart from the assumption made in~\cite{NC}.  In practice, one may characterize the probe states via quantum state tomography (QST)~\cite{banaszek99,james,3,4,altepeter}.  By characterizing the input probe states with QST, the quantum process may be more accurately assessed.  While QST can itself introduce errors, a well calibrated QST system minimizes such errors.  

After the probes emerge from the DUT, they are measured, resulting in a total of eight quantum states (four input and four output), which are measured by QST.   From the measured density matrices, the mapping of the DUT is computed.  The map is normally described as a 4 x 4 process matrix operator ($\chi$).


\begin{figure}
\centering
\scalebox{0.60}{\includegraphics{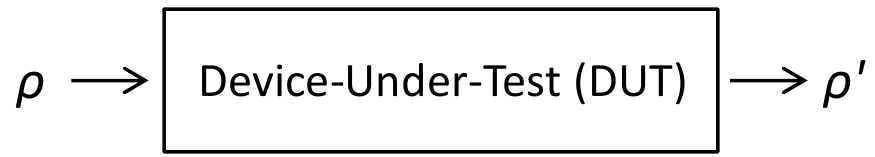}}
\caption{The input state $\rho$ changes to $\rho'$ upon traversal through a device-under-test (DUT). \label{fig:one}}
\end{figure}

Fig.~\ref{fig:one} shows a DUT upon which qubits impinge corresponding to a quantum state described by the density matrix $\rho$. The output qubits' density matrix is denoted by $\rho'$.  Ordinarily QST produces normailized states; however the measurements contain additional information on the loss. To use the loss information, the density matrix of the output state $\rho'$ includes a scaling factor ($\leq 1$) to account for any loss of qubits as they traverse the DUT, that is to say, they are not normalized to the output qubit flux but to the input qubit flux. In other words, while $Tr(\rho)=1$ always, $Tr(\rho')\leq 1$, with the inequality holding for a  non-trace-preserving process.  For example,
\begin{equation}
\rho'=
\left[
\begin{matrix}
1/2&0\\
0&0
\end{matrix}
\right]
=1/2 \left[
\begin{matrix}
1&0\\
0&0
\end{matrix}
\right]
\end{equation}
has trace equal to $\frac{1}{2}$ and may be interpreted as an ensemble of qubits in the pure state $|0\rangle$, reduced in quantity by 50\%.


The output state in Figure 1, which is in general different from the input state due to the action of DUT, can then  be written as
\begin{equation}
\rho'=\boldsymbol\epsilon(\rho),
\end{equation}
where $\boldsymbol\epsilon$ is an operator representing the effect of the DUT on the input state. This can be further expanded as \cite{NC}
\begin{equation}
\boldsymbol\epsilon(\rho) =\sum_iE_i\rho E_i^{\dagger},
\end{equation}
where $E_i$'s comprise a set of at most four operators describing the effect of DUT.  Now these operational elements can be expressed in terms of a \emph{ fixed set of basis operators}, $\tilde{E_k}, k=1,2,..4$, i.e., we can write 
\begin{equation}
E_i=\sum_{m=1}^{4}e_{im}\tilde{E}_m,
\end{equation}
As a result,
\begin{equation}
\rho'=\sum_{mn}\tilde{E}_m\rho\tilde{E}_n^{\dagger}\chi_{mn},
\end{equation}
where 
\begin{equation}
\chi_{mn}=\sum_i e_{im} e_{in}^{*}.
\end{equation}
Since indices $m$ and $n$ each run from 1 through 4, $\chi_{mn}$ is a 4 x 4 matrix, called the\emph{ process matrix}.
This matrix is  Hermitian ($\chi^\dagger=\chi$). Therefore, it has at most $4^2=16$ independent parameters.  Additionally, it is non-negative definite, i.e., its eigenvalues are zero or greater.

Now, invoking the fact that for a trace-preserving process,  $Tr(\rho')=1$, one obtains from Eq.~5  
\begin{equation}
\sum_{mn}\chi_{mn}\tilde{E}_n^\dagger\tilde{E_m}=\mathbbmsl{I}.
\end{equation}
These are, in effect, four constraints on the elements, $\chi_{mn}$. These constraints then reduce the number of independent parameters of the $\chi$ matrix from 16 to 12. In general, including nontrace-preserving processes~\cite{NC,2}, 
\begin{equation}
P\leq \mathbbmsl{I},
\end{equation}
where 
\begin{equation}
P=\sum_i E^\dagger_iE_i=\sum_{mn}\chi_{mn}\tilde{E}_n^\dagger\tilde{E_m}.
\end{equation}
Note that matrix $P$ is Hermitian.

 In what follows, we choose $\tilde{E}_i=\sigma_i$, where $\sigma_1=\mathbbmsl{I}$, $\sigma_2=\sigma_x$, $\sigma_3=\sigma_y$, and $\sigma_4=\sigma_z$, i.e., the identity matrix and the three Pauli spin matrices form the set of basis operators (this is a common choice in quantum computing).  It can be shown that for this \emph{fixed} set of basis of operators $Tr(\chi)=Tr(P)/2$, 
which then equals one for a trace-preserving process because in that case, $P=\mathbbmsl{I}$. Note that the trace of the $\chi$ matrix may not always equal one as it depends on the choice of basis operators as well as the trace-preserving property of the quantum process.

Now Eq.~8 implies that the eigenvalues of the P matrix (defined in Eq.~9) are each less than or equal to one.  For the choice $\tilde{E}_i=\sigma_i$, we find
\begin{widetext}
\begin{equation}
Tr(\chi) +2\sqrt{(Im(\chi_{34})+Re(\chi_{12}))^2+(Im(\chi_{24})-Re(\chi_{13}))^2+(Im(\chi_{23})+Re(\chi_{14}))^2}\leq 1,
\end{equation}
\begin{equation}
Tr(\chi) -2\sqrt{(Im(\chi_{34})+Re(\chi_{12}))^2+(Im(\chi_{24})-Re(\chi_{13}))^2+(Im(\chi_{23})+Re(\chi_{14}))^2}\leq 1.
\end{equation}
\end{widetext}
The left hand side of Eqs.~10 and 11 are the expressions for the eigenvalues of the P matrix in terms of the $\chi$ matrix elements.  $Tr(\chi)=\chi_{11}+\chi_{22}+\chi_{33}+\chi_{44}$. When the process is trace-preserving, the equality  in Eqs.~10 and 11 holds, which then requires that all three terms under the radical sign be individually equal to zero because $Tr(\chi)=1$.  In other words,  constraints  i) $Im(\chi_{34})=-Re(\chi_{12})$, ii) $Im(\chi_{24})=Re(\chi_{13})$, and iii) $Im(\chi_{23})=-Re(\chi_{14})$ must also hold in any numerical fit to the experimental data to yield a physical $\chi$ matrix.  These three  constraints, along with the well-known constraint, $Tr(\chi)=1$, comprise the four constraints of Eq. 7  \footnote{ Note that these explicit forms of the four constraints could also have been derived  directly by solving the four equations that Eq. 7 comprises.}. To our knowledge, this complete set of constraints  in the above simple, explicit forms has not been cited or discussed in the past. Rather, sets of equations of the form, Eq. 7, seem to have  been employed directly as constraints  in  numerical optimization procedures, using Lagrange multipliers, to obtain a fitted physical process matrix from  experimental data (see, e.g., \cite{Obrien,2}). Knowledge of this full set of constraints, especially of the three additional explicit forms expressing relationships among the off-diagonal elements, provides additional insight into the structure of a process matrix.  These explicit forms of the constraint equations can then serve as a useful tool in understanding and interpreting the experimental tomographic data. 

Along with being nonnegative Hermitian, the $\chi$ matrix must, in general, satisfy the above two constraints.  Eq. 10, however, is a much stricter constraint than Eq. 11 (assuming a positive sign for the term involving the radical sign), which implies that the latter is redundant. In other words, Eq. 10 suffices.   This is to be contrasted with the less stringent result $Tr(\chi) \leq 1 ~$\footnote{This result is also obtained by simply adding  Eqs. 10 and 11.},  which is normally quoted in literature (see, e.g.,~\cite{PhysRevLett.97.170501, PhysRevA.80.042103}). Thus, in general, for a quantum process known to be nontrace-preserving like the polarizer (where $Tr(\chi) =1/2$ ideally), or for a process suspected to be not strictly trace-preserving like a  quantum gate with leakage errors, or simply for a DUT whose behavior is not known a priori, the general constraint, Eq. 10, must hold (or be explicitly invoked) in any fit to the experimental data.



Recall that the process matrix for a trace-preserving process has 12 independent parameters, so 12 different measurements on the output quantum states are required. For non-trace preserving processes, one must also include the effect of attenuation for each of the outputs giving rise to four additional parameters for a total of 16.  This is clear from considering the characterization of each of the four output states by quantum state tomography.  For the purpose of discussion, let us consider optical polarization qubits.  In this case, a common choice is to prepare probe states of horizontal ($|0\rangle$), vertical ($|1\rangle$), diagonal ($[|0\rangle+|1\rangle]/\sqrt{2}$), and circular polarizations $[|0\rangle+i|1\rangle]/\sqrt{2}$).  QST of the single qubit requires four independent measurements (corresponding to the four classical Stokes parameters for each of the four output states). Thus a total of 16 measurements are needed to characterize the states output from the DUT.

As is well known, in practice, due to experimental errors, the four initial states  may not be prepared exactly as indicated.  For example, the input horizontal may not exactly correspond to a pure $|0\rangle$ state, but in fact may be an admixture of $|0\rangle$ and $|1\rangle$ states.  Such a mixture could either be coherent corresponding to an angular deviation from true horizontal, or it could be incoherent, corresponding to a slight depolarization error.  Thus quantum state tomography should be performed on the input states as well as the output states.

We now describe the procedure for the construction of the experimental $\chi$ matrix using measured input and output states. It follows the procedure given in ~\cite{NC}, but is modified by the fact that the input states are measured and the process may be nontrace-preserving.  Let $\rho_j (j=1, 2, 3,$ and $4)$ denote the four experimentally prepared and measured input states.  We now write
\begin{equation}
\rho_j=\sum_ir_{ji}\sigma_i.
\end{equation}
The $r_{ji}$ coefficients are then easily determined from Eq.~12.  In this way, each measured input state is decomposed into the basis we have chosen and will be included via $r_{ji}$s  in the calculation of the QPT below.

The measured output states, expressed in terms of the fitted density matrices $\rho''_j$, are first scaled to account for the possibility of a nontrace-preserving process (see the discussion pertaining to Eq.~1):
\begin{equation}
\rho'_j=\rho''_j \frac{I^{(out)}_j}{I^{(in)}_j},
\end{equation}
where $I$ denotes the intensity of the qubit flux, and  the superscripts $(out)$ and $(in)$ refer to the output and input states, respectively. In the quantum photon counting case, $I$ is replaced with $N$, where $N$ denotes the number of photons counted per constant measurement interval. As in Eq.~12, we now write 
\begin{equation}
\rho'_j=\sum_k \lambda_{jk}\sigma_k,
\end{equation}
where the coefficients $\lambda_{jk}$ are determined from the above equation.
Now inserting $\tilde{E}_i=\sigma_i$ (our choice of basis operators) in Eq.~4, we have
\begin{equation}
\rho'_j=
\sum_{mn}\sigma_m\rho_j\sigma_n^\dagger \chi_{mn}.
\end{equation}
Substituting Eqs.~12 and 14 into Eq.~15, we obtain~\footnote{This expression and thus the definition of the $\beta$ coefficients (see Eq.~17) are slightly different from the ones  in \cite{NC}.}
\begin{equation}
\sum_k\lambda_{jk}\sigma_k=\sum_{mn}\sigma_m\sum_i r_{ji}\sigma_i\sigma_n\chi_{mn}.
\end{equation}
We now find we can write
\begin{equation}
\sum_ir_{ji}\sigma_m\sigma_i\sigma_n=\beta^{mn}_{jk}\sigma_k,
\end{equation}
since $\sigma_k$'s form a complete set; $\beta^{mn}_{jk}$ are complex coefficients. Substituting Eq.~17 in Eq.~16, and simplifying, we obtain
\begin{equation}
\lambda_{jk}=\sum_{mn}\beta^{mn}_{jk}\chi_{mn}.
\end{equation}
If we now regard $\beta^{mn}_{jk}$ as forming a 16x16 matrix, whose columns are indexed by $mn$ and rows by $jk$, then the above equation can be rewritten in a compact form as
\begin{equation}
\vec{\lambda}=\beta\vec{\chi},
\end{equation}
where 
$\vec{\lambda}$ and $\vec{\chi}$ are 16 x 1 column vectors whose rows are indexed by $jk$ and $mn$, respectively. Inverting the above equation,
we obtain
\begin{equation}
\vec{\chi}=\kappa\vec{\lambda},
\end{equation}
where $\kappa=\beta^{-1}$. Alternatively, we can write
\begin{equation}
\chi_{mn}=\sum_{jk}\kappa^{jk}_{mn}\lambda_{jk}.
\end{equation}
Having obtained $\chi$ experimentally via the determination of the $r_{ji}$  parameters from Eq.~12, the $\lambda$ parameters from Eq.~14 and the $\beta$ matrix from Eq.~17, it is critical to enforce the following required properties of the process matrix when fitting the experimentally determined process matrix (Eq. 21): 1) $\chi$  is Hermitian, 2) it is nonnegative, and  3) its trace satisfies Eq. 10. 

The above requirements on the experimentally determined process matrix can be enforced via conveniently available convex optimization software packages.  After one has gone through the above ``correction'' procedure on the experimentally determined $\chi$ matrix, one can extract the operation elements, $E_i$, characteristic of the quantum device as shown in ~\cite{NC}.

We have revisited the theoretical aspects of single qubit quantum process tomography to determine the  behavior of  a quantum device. More specifically, we have reexamined the well-known constraints for the process matrix, and recast them into more insightful forms.  In the case of a trace-preserving process, specific relationships among the various elements of the process matrix emerge that then shed light on its basic  generic structure. Knowledge of these new constraint relationships  permit an enhanced understanding of the interpretation and analysis of the experimental data. Moreover, we point out a tighter bound on the trace of the process matrix for a nontrace-preserving process. Implementing the new, simplified form of constraints also serves as an alternative way to numerically fit the experimental data to obtain an optimal, physical process matrix.   Additionally, for completeness, we have described how to compute the process matrix with experimentally measured input states, which goes beyond the general assumption of perfect input state preparation within the framework of quantum process tomography (see, e.g., \cite{NC}).
\bibliography{bib}

\end{document}